\documentclass{aa}
\usepackage{epsfig}

\begin{document}

\title{Possible geometries of afterglow generation in the gamma-ray burst GRB990705 }
\titlerunning{Afterglows of GRBs with X-ray features}

\author{X. Y. Wang,  Z. G. Dai  and T. Lu}
     \institute{Department of Astronomy,Nanjing University, Nanjing 210093, P.R.China\\
              xywang@nju.edu.cn; daizigao@public1.ptt.js.cn; tlu@nju.edu.cn
              }

   \date{Received / Accepted}

\abstract{The absorption feature detected in the prompt X-ray
emission of GRB990705 has important consequences for its
circum-burst environment and therefore on its afterglow. Here we
investigate whether the circum-burst environment constrained by
the absorption feature could be consistent with the observed
$H$-band afterglow, which exhibits an earlier power law decay
($F\propto t^{-1.68}$) but a much faster decay ($\alpha>2.6$;
$F\propto t^{-\alpha}$) about one day after the burst. Two
possible geometries of the afterglow-emitting regions are
suggested: 1) afterglow emission  produced by the impact of the
fireball on the surrounding torus,  which serves as the absorbing
material of the X-ray feature, as would be expected in the models
involving that a supernova explosion precedes the gamma-ray burst
by some time; 2)afterglow {emission} produced in the dense
circum-burst medium inside the torus. In case 1), the faster decay
at the later time is attributed to the disappearance of the shock
due to the counter-pressure in the hot torus illuminated by the
burst and afterglow photons. For case 2), the circum-burst medium
density is found to be very high ( $n\ga 10^4-10^5~ {\rm cm^{-3}}$
) if the emitting plasma is a jet or even higher if it is
spherical. Future better observations of afterglows of GRBs that
have absorption features might make it possible to make a more
definite choice between these two scenarios.
 \keywords gamma rays: bursts---line:
formation---radiation mechanism: nonthermal
               }

\titlerunning{Afterglows of GRBs with X-ray features}
   \maketitle
%
%_____________________________________________________

\section{Introduction}
There is increasing observational evidence favoring the existence
of absorption and emission lines in the X-ray spectra of gamma-ray
bursts (GRBs) and their afterglows. Emission or absorption
features can provide a fundamental tool for studying  the close
environment of GRBs (e.g. M\'{e}sz\'{a}ros \& Rees 1998; Lazzati
et al. 1999, 2002; B\"{o}ttcher \& Fryer 2001). To date, five
bursts have shown evidence for  iron or lighter element emission
lines during the X-ray afterglow (GRB970508, Piro et al. 1998;
GRB970828,  Yoshida et al. 1999; GRB991216, Piro et al. 2000;
GRB000214,  Antonelli et al. 2000; GRB011211, Reeves et al. 2002)
and one (GRB990705; Amati et al. 2000; hereafter A2000) displays a
transient absorption feature at 3.8 KeV during the burst itself.

A few models for emission lines in the X-ray afterglows have been
suggested (see Piro 2002 for a review ), including ``distant
reprocessor scenario" and ``nearby reprocessor scenario". In the
former, the line-emitting gas is located at $R\ga 10^{15}~{\rm
cm}$ with the line variability time corresponding to the light
travel time between GRB and the reprocessor (Lazzati et al. 1999;
Piro 2000; Weth et al. 2000). This scenario needs the presence of
an iron-rich dense medium with iron mass $M_{\rm Fe}\ga 0.01
M_\odot$. The most straightforward picture is the one in which  a
SN-like explosion occurs some time before the formation of the
GRB. The GRB may be produced by the collapsing of the
rotationally-supported newborn massive neutron star to a black
hole (Vietri \& Stella 1998), or the phase transition to a strange
star (Wang et al. 2000a) . In the latter scenario, the line
emission is attributed to the interaction of a long-lasting
relativistic outflow from the central engine with the massive star
progenitor stellar envelope at distances $R\la10^{13}~{\rm cm}$
 (M\'{e}sz\'{a}ros \& Rees 2000; Rees \& M\'{e}sz\'{a}ros 2000).

While different scenarios have been suggested to explain the
emission line, the properties of the {transient} absorption
feature, as in GRB990705, strongly point to a unique {circum-burst
environment (Lazzati et al. 2001; B\"{o}ttcher et al. 2002),
i.e.}, 1) iron-rich absorbing matter of a few solar masses (such
as the young supernova {remnant} shell ) lies between $10^{16}$
and $10^{18}~$cm from the burst site; 2) the absorbing matter is
located in the line of sight between the observer and the burster.

GRB990705 has a duration of $\sim42 ~{\rm s }$ in the Gamma-Ray
Burst Monitor (GRBM) and fluence $(9.3\pm 0.2)\times10^{-5}~{\rm
erg~ cm^{-2}}$ in the $2-700 {\rm keV}$ band (A2000). During the
prompt phase, it shows an absorption feature at 3.8 keV and an
equivalent hydrogen column density, which disappears 13 s after
the burst onset (A2000). This absorption feature was explained by
A2000 as being due to an edge produced by neutral iron redshifted
to $3.8\pm0.3~{\rm keV}$; the corresponding redshift is
$0.86\pm0.17$. Optical spectroscopy of the host galaxy gives a
redshift $z=0.8435$ (Andersen et al. 2002), consistent with the
inferred value from the X-ray feature. This straightforward
interpretation was, however, questioned by Lazzati et al. (2001)
as it requires a vast amount of iron {\footnote { The required
total mass of iron is $35 f M_\odot$ ( see Eq.(5) in Lazzati et
al. 2001), where $f$ is the covering factor of the absorbing
material surrounding the burst.}}  in the close vicinity of the
burster. Lazzati et al. (2001) further suggested an alternative
scenario in which the feature is produced by resonant scattering
from hydrogen-like iron broadened by a range of outflow
velocities. In this scenario, the radius of the SN shell is fixed
by the requirement that the heating timescale of the electrons in
the absorbing matter is $\sim 10~{\rm s}$, i.e. $R_s\sim
(2-3)\times10^{16}~{\rm cm}$. {Our following work is based on this
scenario.}

A fading X-ray afterglow of GRB990705 was detected by the Narrow
Field Instruments of {\it BeppoSAX} 11 hours after the trigger,
but the statistics are not sufficient to draw a detailed
conclusion on the decaying law (A2000). Masetti et al. (2000)
report having detected  the counterpart of this burst twice in the
near-infrared $H$ band and only once in the optical $V$ band, from
a few hours to $\sim1$ day after the GRB trigger. The first two
$H$-band measurements define a power-law decay with index
$\alpha=1.68\pm0.10$ ($F\propto{t^{-\alpha}})$, but a third
attempt to detect the source gave an upper limit, implying a much
faster decay. No radio afterglow was detected (Subrahmanyan et al.
1999; Hurley et al. 1999).

For the afterglows with X-ray {\em emission }lines, the
line-emitting gas  could lie outside of the line of sight of the
burst and therefore has no direct relation with the afterglow
radiation. However,  for afterglow with X-ray {\em absorption }
features, the absorbing matter (SN shell) should have a direct
consequence on the afterglow radiation, because it  must lie in
the line of sight of the burst. So, an examination of the
self-consistency between the power-law afterglow and the X-ray
absorption feature is quite necessary.

\section{Afterglow models for GRB990705}

We here investigate the afterglow behavior of GRBs assuming the
supranova-like scenario  ( Vietri \& stella 1998; Wang et al.
2000a) where a thick torus of matter (i.e. the supernova remnant
shell ) lies, in the line of sight of the burst, at a radius $R_s$
from the burst center with a width $\Delta R_s$ and particle
density $n_s$.  We attempt to fit the $H$-band afterglow of
GRB990705, as a representative case. For uniform circum-burst
medium, the deceleration radius, at which the energy of the hot,
swept-up external medium by the blast wave equals that in the
original explosion, of
 the GRB { relativistic shell} is (e.g. Piran 1999)
\begin{equation}
R_d=(\frac{3E}{4\pi \eta^2 n m_p c^2})^{1/3}=6\times10^{16} ~{\rm
cm}~E_{53}^{1/3}n_0^{1/3}\eta_{300}^{-2/3}
\end{equation}
where $E=10^{53}\,E_{53}\,{\rm erg}$ is the  shell isotropic
kinetic energy, $n=10^0 ~{\rm cm^{-3}}$ is the particle density of
the circum-burst medium, and $\eta=300\,\eta_{300}$ is the initial
Lorentz factor of the  shell. According to whether $R_d\gg R_s$ or
$R_d\ll R_s$, there are  two  possible locations of the
afterglow-emitting regions: one is in the torus on which the
fireball impacts (case I) and the other is in the circum-burst
medium inside the torus (case II).

\subsection{case I: jet-torus interaction model}
We assume that the torus has a width $\Delta R_s$, density
$n_s=M/4\pi R^2 m_p$ and scattering optical depth $\tau_T=\sigma_T
n\Delta R_s$. $\tau_T\la 1$ must be satisfied to maintain the
flickering behavior of the burst. Values consistent with this
could be a few solar masses located at
$R_s\sim(2-3)\times10^{16}~{\rm cm}$, which gives
$\tau_T=0.67(M/10M_\odot)(R_s/3\times10^{16}{\rm cm})^{-2}$ and a
particle density $n_s=10^9(M/10M_\odot)(R_s/3\times10^{16}{\rm
cm})^{-2}(\Delta R_s/10^{15}{\rm cm})^{-1}$.

The torus will be hit by the the fireball shell a few seconds
($\delta t\sim R_s/2\eta^2 c=2~ s~ R_{s,16}\eta_{300}^{-2}$, where
$R_s=10^{16}R_{s,16}~{\rm cm}$) after it is reached by the burst
proper. The impact process has been described in Vietri et al.
(1999), where the authors attempt to interpret the anomalous X-ray
afterglow of GRB970508 and GRB970828. The impact of the fireball
on the torus will generate a forward shock propagating into the
torus, and a reverse one moving into the fireball shell. They
predicted, during the impacting, a secondary burst from the
reverse shock and a very short-lived forward shock for  GRB970508.
However, we will show below that for GRB990705, which has a much
larger shock energy $E$, the forward shock could last few days (
especially for the lower estimated value for the torus temperature
given by Paerels et al. (2000), see Eq. (7)) , giving rise to an
early power-law fading afterglow as seen in GRB990705. The
disappearance of this forward shock  may just account for the
observed faster decline at the late time.

 When the rest mass-energy of the swept-up material  equals
the shock energy,
 the forward shock will be slowed down to
non-relativistic speeds, which occurs after the shock has
propagated a quite short distance $d$ in the torus {\footnote{The
denominator of the formula of $d$ in Vietri et al.(1999) has an
extra ``$\eta$", so they got a much lower value for $d$. }}, where
\begin{equation}
d=\frac{E}{4\pi R_s^2 n_s m_p c^2}=5\times10^{12}~{\rm cm}~ E_{53}
n_{s,10}^{-1}R_{s,16}^{-2},
\end{equation}
and the corresponding time  is
\begin{equation}
t_{nr}=\frac{d}{c}=160~{\rm s}~E_{53} n_{s,10}^{-1}R_{s,16}^{-2}.
\end{equation}
For an adiabatic shock, the conservation of energy is as follows
\begin{equation}
E=4\pi R_s^2 x n_s m_p v^2/2={\rm constant}
\end{equation}
where $x$ is the distance that the forward shock have propagated
{\em in the torus }and $v$ is the shock velocity. From this
equation and $t\sim {x}/{v}$ , we get the scaling laws of the
dynamic quantities: $v=c (x/d)^{-1/2}$, $v=c (t/t_{nr})^{-1/3}$
and $x=d(t/t_{nr})^{2/3}$. Please note that these dynamic
relations are different from the usual Sedov-von Neumann-Taylor
solution of a non-relativistic GRB shock (Wijers et al. 1997; Dai
\& Lu 1999; Wang et al. 2000b) because  here the fireball is
decelerated in a dense shell with an almost fixed radius $R_s$.

As the fireball slows down, the ram pressure of the shell
($P=\rho_b  v^2 $ where $\rho_b=E/\eta c^2 4\pi R_s^2 m_p x_b $ is
the shell density) on the external torus matter decreases with
time. The material in the torus is supposed to be brought up to a
temperature $T_s\sim 10^7-10^8~{\rm K}$ by heating/cooling from
the proper burst and its afterglow radiation (Vietri et al. 1999;
Paerels et al. 2000). Thus, at a certain distance $x_b$ , the
strong counter-pressure ($\sim n_s k T_s$) in the pre-shock torus
equals the ram pressure and  begins to damp down the forward
shock. We expect that the forward shock emission decays
exponentially with time since then. Equating $\rho_b v^2$ with
$n_s k T_s$ gives
\begin{equation} \begin{array}{ll}
x_b\simeq (\frac{E d}{8\pi \eta R_s^2 n_s k T_s})^{1/2} \\
=2\times10^{14} ~{\rm
cm}~E_{53}R_{s,16}^{-2}\eta_{300}^{-1/2}n_{s,10}^{-1}T_{s,7}^{-1/2}.
\end{array}
\end{equation}
The shock velocity at $x_b$ is
\begin{equation}
v_b=c(x_b/d)^{1/2}=0.16c ~ \eta_{300}^{1/4}T_{s,7}^{1/4}.
\end{equation}
So,  the characteristic time when the forward shock vanishes is
\begin{equation}
t_b\sim x_b/v_b=4\times10^{4} ~{\rm
s}~E_{53}R_{s,16}^{-2}\eta_{300}^{-3/4}n_{s,10}^{-1}T_{s,7}^{-3/4}
\end{equation}
after the burst (note that here $\delta t$ and $t_{nr}$, compared
to $t_b$, are both negligible ).

Up to now, we have assumed that   the radial time scale of the
fireball shell is relevant to the dynamic time scale. This
requires that the angular spreading timescale does not dominate
the radial time scale, i.e. $R_s\theta_j^2/2c\la{x/v}$, where
$\theta_j$ is the opening angle of the fireball shell, which means
that actually the outflow is a jet. The first measurement of the
$H$-band afterglow is at $\sim 4$ hours after the burst, so
$\theta_j\la 0.3 R_{s,16}^{-1/2}$. Actually, a mildly collimated
outflow is quite plausible  considering the large isotropic
gamma-ray energy of this burst. Please note that, in the jet-torus
interaction model, sideways expansion of the jet in the torus
cannot change the opening angle significantly as the sideways
expansion length is much smaller than the radius $R_s$, i.e.
$\theta_j=\theta_0+c_s t/(R_s+vt)\simeq \theta_0$, where $c_s$ is
the sound velocity in the torus.

Now we investigate the fading behavior of the afterglow as the
non-relativistic forward shock slows down in the torus. During
this phase, the typical electron Lorentz factor is
\begin{equation} \begin{array}{ll}
\gamma_m=\epsilon_e\frac{(p-2)}{(p-1)}\frac{m_p}{m_e}\frac{v^2}{2c^2}\\
=60\frac{(p-2)}{(p-1)}\epsilon_{e,0.5}E_{53}^{2/3}n_{s,10}^{-2/3}R_{s,10}^{-4/3}t_{1h}^{-2/3},
\end{array}
\end{equation}
where $\epsilon_e\equiv0.5\epsilon_{e,0.5}$ is the fraction of the
shock energy carried by the electrons and $t_{1h}$ is the
observing time in units of one hour. The post-shock magnetic field
strength is
\begin{equation}\begin{array}{ll}
B=\sqrt{8\pi\epsilon_B (4n_sm_pv^2/2)} \\
=100~{\rm
G}~\epsilon_{B,-4}^{1/2}n_{s,10}^{1/6}E_{53}^{1/3}R_{s,16}^{-2/3}t_{1h}^{-1/3},
\end{array}
\end{equation}
where $\epsilon_B\equiv10^{-4}\epsilon_{B,-4}$ is the fraction of
the shock energy carried by the magnetic field. Thus we obtain the
synchrotron peak frequency
\begin{equation}\begin{array}{ll}
\nu_m=\frac{\gamma_m^2q_eB}{2\pi m_e c} \\
= 10^{12}~{\rm Hz}~(\frac{p-2}{p-1})^2
\epsilon_{e,0.5}^2\epsilon_{B,-4}^{1/2}E_{53}^{5/3}n_{s,10}^{-7/6}R_{s,16}^{-10/3}t_{1h}^{-5/3}
\end{array}
\end{equation}
where $q_e$ is the electron charge, and the cooling frequency
\begin{equation}
\nu_c=6\times10^{10}~{\rm
Hz}~\epsilon_{B,-4}^{-3/2}n_{s,10}^{-1/2}E_{53}^{-1}R_{s,16}^2
t_{1h}^{-1} .
\end {equation}
The peak flux is
\begin{equation}
F_{\nu_m}= \frac{1}{4\pi d_L^2}\frac{q_e^3}{m_e c^2}N_e B \propto
t^{1/3},
\end{equation}
where $N_e=4\pi R_s^2 x n_s \theta_j^2/2$ is the total number of
electrons swept-up  by the forward shock, $\theta_j$ is the
opening angle of the jet and $d_L$ is the luminosity distance of
the burst. According to these relations, we further derive the
spectrum and the light curve:

\begin{equation}
 F_{\nu} =  \left\{ \begin{array}{llll}
  (\nu/\nu_m)^{-(p-1)/2}F_{\nu_m}    \\
 \,\,\, \propto \nu^{-(p-1)/2}t^{-5p/6+7/6}      & {\rm if}\,\, \nu_c>\nu>\nu_m                \\
  (\nu_c/\nu_m)^{-(p-1)/2}(\nu/\nu_c)^{-p/2}F_{\nu_m}       \\
 \,\,\, \propto \nu^{-p/2} t^{-5p/6+2/3} & {\rm if} \,\, \nu>\nu_c>\nu_m

                                             \end{array} \right. \;,
\end{equation}
Thus the $H$-band decay index of GRB990705 before 1 day can be
reproduced if $p\simeq 2.8$ and $\nu_{H}>\nu_c>\nu_m$. In Fig.1,
we give an analytic fitting to the $H$-band afterglow. As
discussed in sect. 1, the radius of the torus is almost fixed by
the requirement that the heating timescale of the electrons in the
torus is $\sim10~{\rm s}$ and we set $R_s=3\times10^{16}~{\rm
cm}$. For a SN shell of $10M_\odot$ mass with a typical width
$\Delta R=10^{15}~{\rm cm}$ at this radius, its particle density
is $n_s=10^9~{\rm cm^{-3}}$. We also chose the shock energy to be
$E=5\times10^{53}~{\rm erg}$, according to the isotropic gamma-ray
energy of GRB990705 and a reasonable gamma-ray production
efficiency of GRBs.  The free parameters in this fitting are the
two unknown energy equipartition factors $\epsilon_e$ and
$\epsilon_B$, and the opening angle of the jet $\theta_j$. We find
that the following values for these parameters are consistent with
the observations: $\epsilon_e=0.5$, $\epsilon_B=10^{-5}$ and
$\theta_j=0.2$. The solid line in Fig.1 represents the power-law
 decay of the afterglow as the forward shock slows down in the torus and
  the thick dotted line represents the later exponential decay ($F_\nu \propto {\rm exp(-t)}$)
  of this shock due to the counter-pressure in the hot torus. With these parameters, the synchrotron self-absorption frequency
scales with time as
\begin{equation}
\nu_a=930~{\rm GHz}~(t/1 \, d)^{-(6-5p)/3(p+4)}.
\end{equation}
Such a large synchrotron self-absorption frequency is consistent
with the non-detection of the radio afterglow.
\begin{figure}
\centerline{\hbox{\psfig{figure=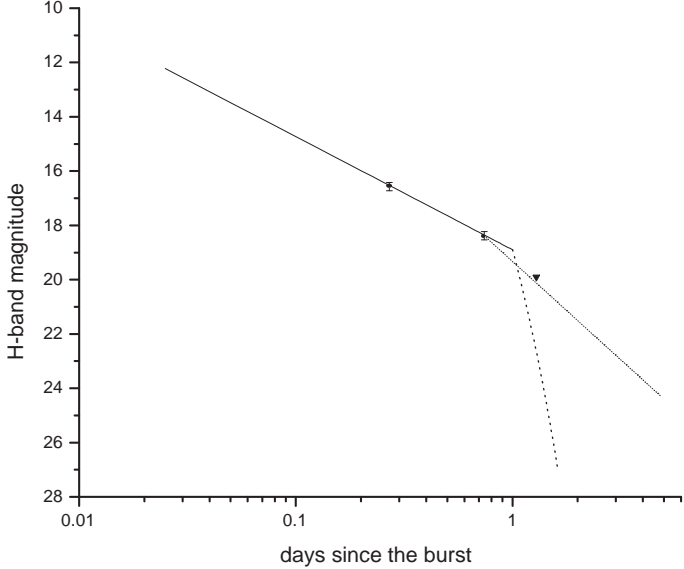,width=4.0in,angle=0}}}
\caption {An analytic fitting of the $H$-band afterglow of
GRB990705 in terms of the jet-torus interaction model  (see text
for details). Detections and upper limits for the non-detections,
 taken from  {  Masetti et al.} (2000),
 are indicated by the filled circles with error bars ($3\sigma$) and arrows, respectively. The solid line represents the power-law
 decay of the afterglow as the forward shock slows down in the torus and
  the  thick dotted line represents the late exponential decay of this shock due to the effect of the counter-pressure in the hot torus.
  The thin dotted line  represents the
late power-law decay $F_\nu \propto t^{-p}$ due to jet behavior.
See the text for the parameters used in this fitting. }
\end{figure}

The bremsstrahlung cooling time of the torus of density
$n\sim10^9~ {\rm cm^{-3}}$ is given by
\begin{equation}
t_{br}=7\times10^{5}~ {\rm s}~n_9^{-1}T_{s,7}^{1/2},
\end{equation}
so the hot torus does not cool significantly during the phase of
the interaction between the jet and the torus.

\subsection{Case II: jet in a dense circum-burst medium }
The steepness of the light curve decay could  also be produced by
a  beamed outflow (e.g. Rhoads 1999; Sari et al. 1999). The beam
reduces the energy budget, alleviating the ``energy crisis" of
GRBs. Assuming that a break due to jet sideways spreading occurs
in the $H$-band light curve of  GRB990705 about one day after the
burst, the early time slope $\alpha\simeq 1.68$ and the later one
$\alpha'>2.6$ (  based on the second $H$-band detection and the
third $H$-band upper limit ) would be consistent with $p\sim2.9$.
The thin dotted line in Fig.1 represents this later power-law
decay $F_\nu \propto t^{-p}$ due to jet sideways expansion
behavior. The sideways expansion of the jet makes its bulk Lorentz
factor $\Gamma$ slow down exponentially with radius after a
characteristic value $\theta_j^{-1}$. Afterwards, $\Gamma\propto
{\rm exp}(-r/R_b)$, where $R_b$ is the shock radius at the time
$\Gamma=\theta_j^{-1}$. For a uniform circum-burst medium, we have
$\Gamma=(17E/1024\pi n m_p c^5 t^3)^{1/8}$ (Sari et al. 1998), and
\begin{equation}
R_b=(\frac{17E_0}{8\pi n m_p c^2})^{1/3}=7\times10^{17}~{\rm
cm}~E_{0,51}^{1/3}n_0^{-1/3}
\end{equation}
where $E_0$ is the actual energy of the jet, $E_0=E_{\rm
iso}\theta_j^2/2$. If the early power-law decaying afterglow is
assumed to be produced by the deceleration of the jet in the
circum-burst medium before it hits the surrounding  torus, we
require $R_b<R_s\sim3\times10^{16}~{\rm cm}$. This means that the
circum-burst medium has a number density $n\ga10^4-10^5 ~{\rm
cm^{-3}}$, even if the actual energy of the burst is only $E_0\sim
{\rm a\, few }\times10^{51}~{\rm erg}$ as found by Frail et al.
(2001). Frail et al. (2001) have inferred the jet opening angle
$\theta_j\simeq 0.054$ from the light curve break  time, assuming
an interstellar medium of density $n=0.1~{\rm cm^{-3}}$. A much
larger circum-burst medium density leads to  an energy reservoir
an order of magnitude larger than what estimated by Frail et al.
(2001), as $E_0\propto \theta_j^2\propto n^{1/4}$. A much larger
density than that of a typical interstellar medium is also
suggested by Ghisellini et al. (2002) from the point of view of
constraining the total energy reservoir of GRB991216  with
emission line luminosity.

Afterglow light curve breaks can also be produced by spherical
fireball expansion which undergoes a transition from a
relativistic phase to a non-relativistic one (Wijers, Rees \&
M\'{e}sz\'{a}ros 1997; Dai \& Lu 1999; Livio \& Waxman 2000;). The
power-law decay indices before and after the break are consistent
with $p\sim3.2$ if the $H$-band frequency is located between the
characteristic break frequency and the cooling break frequency
during the first day after the burst (see Eqs.(5) and (6) of Dai
\& Lu 1999 ). This scenario also requires that at least the Sedov
length of the shock $R_{nr}$ is less than the torus radius. As
\begin{equation}
R_{\rm nr}=(\frac{E_{\rm iso}}{4\pi/3n m_p
c^2})^{1/3}=2.5\times10^{18}~{\rm cm}~E_{{\rm
iso},53}^{1/3}n_0^{-1/3},
\end{equation}
where $E_{iso}$ is the isotropic kinetic energy of burst, it means
$n\ga 10^{6}~{\rm cm^{-3}}$. Such a large number density ($n\ga
10^4-10^6 ~{\rm cm^{-3}}$) is typical of molecular clouds in star
forming regions, independently supporting that long GRBs are
connected with massive progenitors.

{Due to the lack of detection after about one day for GRB990705,
we do not
 know the later behavior of its afterglow, hence we could not tell these scenarios from each other for this
burst.  However, we predict different types of behavior of
afterglows for the different geometries discussed above.} If we
could know the spectra and light curves of the afterglows (for
those GRBs that have absorption features) both before and after
the break in future better observations, we can then have a more
definite conclusion. The difference in the spectrum and light
curve for these different scenarios are summarized in Table 1.
\begin{center}
\begin{table*}[ht!]
\begin{center}
\begin{tabular}{|c||c||c|c|}
\hline & & \multicolumn{2}{|c|}{Case II~~~~~~~~~~~~} \\
& \raisebox{1.5ex}[0pt] {Case I } & jet & relativistic to non-relativistic \\
\hline\hline {$t<t_b,\nu<\nu_c$}&
{$\nu^{-({p-1})/{2}}t^{-({5p-7})/{6}}$} &
$\nu^{-(p-1)/2}t^{-3(p-1)/4}$ & $\nu^{-(p-1)/2}t^{-3(p-1)/4}$
 \\ \hline
$t<t_b,\nu>\nu_c$& $\nu^{-p/2}t^{-(5p-4)/6}$ &$\nu^{-p/2}t^{-(3p-2)/4} $& $\nu^{-p/2}t^{-(3p-2)'4}$ \\
 \hline
$t>t_b,\nu<\nu_c$& $\nu^{-(p-1)/2}{\rm exp}(-t)$ & $\nu^{-(p-1)/2}t^{-p}$ & $\nu^{-(p-1)/2}t^{(21-15p)/10}$ \\
 \hline
$t>t_b,\nu>\nu_c$&$\nu^{-p/2}{\rm exp}(-t)$  & $\nu^{-p/2}t^{-p}$ & $\nu^{-p/2}t^{(4-3p)/2}$ \\
 \hline
\end{tabular}
\end{center}
\par
\label{t:afterglow} \caption{The spectra and light curves for the
different scenarios discussed in the text. $t_b$ is the light
curve break time of the afterglows. The parameter free relation
between the spectral index  and the light curve index can be
derived by eliminating $p$ for each case. }
\end{table*}
\end{center}
\section{Conclusions and discussions}
Emission or absorption features in the X-ray spectrum of GRBs and
their afterglows provide a useful tool  for studying the close
environment of GRBs and thus their possible progenitors. The
absorption feature in the prompt X-ray emission of GRB990705 was
originally interpreted by Amati et al. (2000) to be a
photoionization  K edge of neutral iron. However, this
straightforward explanation is shown by Lazzati et al. (2001) to
 require an improbably large  amount of iron in the
close environment of the burster. Instead, Lazzati et al. (2001)
interpret this as a resonant absorption line broadened by a large
spread of velocities. In this scenario, the disappearance of the
feature 13 s after the burst results from electron heating due to
the illuminating photons and it severely constrains the radius of
the absorbing materials ($R\sim2-3\times10^{16}~{\rm cm}$, see Eq.
(13) of Lazzati et al. 2001). A reasonable scenario for this
requirement is the supranova-like scenarios  ( Vietri \& Stella
1998; Wang et al. 2000a), in which a young supernova remanent  is
located at the close vicinity of the burster. Based on these
studies, in this paper we investigated whether the circum-burst
environment constrained by the absorption feature could be
consistent with the observed afterglows of GRB990705.

We discussed  two  possible locations of the afterglow-emitting
region: one is in the torus where the afterglows are produced by
the impact of the fireball jet on this torus and the other is in
the dense circum-burst medium inside the torus. In the former
scenario, the impact of the fireball on the torus will generate a
forward shock propagating into the torus. This forward shock will
be decelerated by the dense matter in the torus into a
sub-relativistic phase in quite a short time and to a lower and
lower velocity as  time elapses. The heating/cooling processes of
the torus by the burst and afterglow photons may bring its
temperature to $T_s\sim10^7~{\rm K}$. Once  the ram pressure
($\sim \rho_b v^2$ ) of the fireball  falls low enough to be equal
to the thermal counter-pressure ($n_skT_s$) of the hot torus,  the
forward shock
 is damped down very rapidly (Vietri et al. 1999) and the
afterglow emission will cut off accordingly. We found that the
$H$-band afterglow of GRB990705 can be fitted in terms of this
model.

In the latter scenario, as in many other afterglows, the steeping
of light curve decay of GRB990705 one day after the burst is
attributed to the jet evolution in a uniform density medium or a
spherical fireball undergoing a transition to non-relativistic
expansion. The broken power-law decay behavior of the $H$-band
afterglow requires the shock radius at the light curve break time
or at the Sedov phase, respectively, to be smaller than the torus
location. This in turn requires that the circum-burst medium
density must be $n\ga10^4-10^5~{\rm cm^{-3}}$ or $n\ga10^6~{\rm
cm^{-3}}$, respectively. In this scenario, the fireball will also
hit the surrounding torus finally. The abrupt density jump  might
cause a rise and a successive decline in the afterglows (see Dai
\& Lu 2002 for a relativistic case).

A noticeable point relevant to the high density circum-burst
medium is that the true energy reservoir  of GRB990705 may be much
greater than what was estimated by Frail et al. (2001),
$E_\gamma=3.9\times10^{50}~{\rm erg}$, derived from the jet model
by assuming an interstellar medium of density $n=0.1~{\rm
cm^{-3}}$, since the calculated fireball true energy depends on
$\theta_j^2$ which in turn depends on $n^{1/4}$.

In summary, the geometry requirement of  the X-ray absorption
feature of GRB990705 is shown to be also consistent with its
afterglows, although the sparse data of the afterglow makes it
impossible to reach a definite conclusion on the two scenarios.
Future better broad-band observations of the afterglow spectra and
light curves for GRBs that have absorption features could tell
which one is true and thereby provides a more valuable insight
into the environment and the central engine.

\begin{acknowledgements}
 We are grateful to the referee for his constructive and careful
comments. XYW would like to thank Z. Li for valuable discussions.
This work was supported by the National Natural Science Foundation
of China under grants 19973003 and 19825109, and the National 973
project.

\end{acknowledgements}

\end{document}